# View Selection in Semantic Web Databases [*]

François Goasdoué[1]    Konstantinos Karanasos[1]    Julien Leblay[1]    Ioana Manolescu[1]

[1]Leo team, INRIA Saclay and LRI, Université Paris-Sud 11
firstname.lastname@inria.fr

## ABSTRACT

We consider the setting of a Semantic Web database, containing both explicit data encoded in RDF triples, and implicit data, implied by the RDF semantics. Based on a query workload, we address the problem of selecting a set of views to be materialized in the database, minimizing a combination of query processing, view storage, and view maintenance costs. Starting from an existing relational view selection method, we devise new algorithms for recommending view sets, and show that they scale significantly beyond the existing relational ones when adapted to the RDF context. To account for implicit triples in query answers, we propose a novel RDF query reformulation algorithm and an innovative way of incorporating it into view selection in order to avoid a combinatorial explosion in the complexity of the selection process. The interest of our techniques is demonstrated through a set of experiments.

## 1. INTRODUCTION

A key ingredient for the Semantic Web vision [4] is a data format for describing items from the real and digital world in a machine-exploitable way. The W3C's resource description framework (RDF, in short [26]) is a leading candidate for this role.

At a first look, querying RDF resembles querying relational data. Indeed, at the core of the W3C's SPARQL query language for RDF [27] lies conjunctive relational-style querying. There are, however, several important differences in the data model. First, an RDF data set is a single large set of triples, in contrast with the typical relational database featuring many relations with varying numbers of attributes. Second, RDF triples may feature *blank nodes*, standing for unknown constants or URIs; an RDF database may, for instance, state that the *author* of $X$ is *Jane* while the *date* of $X$ is *4/1/2011*, for a given, unknown resource $X$. This contrasts with standard relational databases where all attribute values are either constants or *null*. Finally, in typical relational databases, all data is *explicit*, whereas the semantics of RDF entails a set of *implicit* triples which must be reflected in query answers. One important source of implicit triples follows from the use of an (optional) RDF Schema (or RDFS, in short [26]), to enhance the descriptive power of an RDF data set. For instance, assume the RDF database contains the fact that the *driverLicenseNo* of *John* is *12345*, whereas an RDF Schema states that only a *person* can have a *driverLicenseNo*. Then, the fact that *John* is a *person* is implicitly present in the database, and a query asking for all *person* instances in the database must return *John*.

The complex, graph-structured RDF model is suitable for describing heterogeneous, irregular data. However, it is clearly not a good model for storing the data. Existing RDF platforms therefore assume a simple (application-independent) storage model, complemented by indexes and efficient query evaluation techniques [1, 15, 16, 17, 20, 23], or by RDF materialized views [6, 9]. While indexes or views speed up the evaluation of the fragments of queries matching them, the query processor may still need to access the main RDF database to evaluate the remaining fragments of the queries.

We consider the problem of *choosing a (relational) storage model for an RDF application*. Based on the application workload, we seek a set of views to materialize over the RDF database, such that *all workload queries can be answered based solely on the recommended views, with no need to access the database*. Our goal is to enable three-tier deployment of RDF applications, where clients do not connect directly to the database, but to an application server, which could store only the relevant views; alternatively, if the views are stored at the client, no connection is needed and the application can run off-line, independently from the database server.

RDF datasets can be very different: data may be more or less structured, schemas may be complex, simple, or absent, updates may be rare or frequent. Moreover, RDF applications may differ in the shape, size and similarity of queries, costs of propagating updates to the views etc. To capture this variety, we characterize candidate view sets by a cost function, which combines ($i$) query evaluation costs, ($ii$) view maintenance costs and ($iii$) view storage space. Our contributions are the following:

**1.** This is the first study of RDF materialized view selection supporting the rewriting of all workload queries. We show how to model this as a search problem in a space of states, inspired from a previous work in relational data warehousing [21].
**2.** Implicit triples entailed by the RDF semantics [26] must be reflected in the recommended materialized views, since they may participate to query results. Two methods are currently used to include implicit tuples in query results. *Database saturation* adds them to the database, while *query reformulation* leaves the database intact and modifies queries in order to also capture implicit triples. Our approach requires no special adaptation if applied on a saturated database. For the reformulation scenario, we propose a novel RDF query reformulation algorithm. This algorithm extends the state of the art in query processing in the presence of RDF Schemas [3,

[*]This work has been partially funded by *Agence Nationale de la Recherche*, decision ANR-08-DEFIS-004.

5], and is a contribution applying beyond the context of this work. Moreover, we propose an innovative method of using reformulation (called *post-reformulation*) which enables us to efficiently take into account implicit triples in our view selection approach.

**3.** We consider heuristic search strategies, since the complexity of complete search is extremely high. Existing strategies for relational view selection [21] grow out of memory and fail to produce a solution when the number of atoms in the query workload grows. Since RDF atoms are short (just three attributes), RDF queries are syntactically more complex (they have more atoms) than relational queries retrieving the same information, making this scale problem particularly acute for RDF. We propose a set of new strategies and heuristics which greatly improve the scalability of the search.

**4.** We study the efficiency and effectiveness of the above algorithms, and their improvement over existing similar approaches, through a set of experiments.

This paper is organized as follows. Section 2 formalizes the problem we consider. Section 3 presents the view selection problem as a search problem in a space of candidate states, whereas Section 4 discusses the inclusion of implicit RDF triples in our approach. Section 5 describes the search strategies and heuristics used to navigate in the search space. Section 6 presents our experimental evaluation. Section 7 discusses related works, then we conclude.

## 2. PROBLEM STATEMENT

In accordance with the RDF specification [26], we view an RDF database as a set of $(s, p, o)$ triples, where $s$ is the *subject*, $p$ the *property*, and $o$ the *object*. RDF triples are *well-formed*, that is: subjects can be URIs or *blank nodes*, properties are URIs, while objects can be URIs, blank nodes, or literals (i.e., values). Blank nodes are placeholders for unknown *constants* (URIs or literals); from a database perspective, they can be seen as existential variables in the data. While relational tuples including the $null$ token, commonly used to represent missing information, do not join ($null$ does not satisfy any predicate), RDF triples referring to *the same* blank node may be joined to construct complex results, as exemplified in the Introduction. Due to blank nodes, an RDF database can be seen as an incomplete relational database consisting of a single *triple table* $t(s, p, o)$, under the open-world assumption [2].

To express RDF queries (and views), we consider the basic graph pattern queries of SPARQL [27], represented *wlog* as a special case of conjunctive queries: conjunctions of atoms, the terms of which are either free variables (a.k.a. head variables), existential variables, or constants. We do not use a specific representation for blank nodes in queries, although SPARQL does, because they behave exactly like existential variables.

DEFINITION 2.1 (RDF QUERIES/VIEWS). *An RDF query (or view) is a conjunctive query over the triple table* $t(s, p, o)$.

We consider *wlog* queries *without Cartesian products*, i.e., each triple shares at least one variable (joins at least) with another triple. We represent a query with a Cartesian product by the set of its independent sub-queries. Finally, we assume queries and views are *minimal*, i.e., the only containment mapping from a query (or view) to itself is the identity [7].

As a running example, we use the following query $q_1$, which asks for painters that have painted "Starry Night" and having a child that is also a painter, as well as the paintings of their children:

$q_1(X, Z) :- t(X, hasPainted, starryNight), t(X, isParentOf, Y), t(Y, hasPainted, Z)$

Based on views, one can rewrite the workload queries:

DEFINITION 2.2 (REWRITING). *Let $q$ be an RDF query and $V = \{v_1, v_2, \ldots, v_k\}$ be a set of RDF views. A rewriting of $q$ based on $V$ is a conjunctive query (i) equivalent to $q$ (i.e., on any data set, it yields the same answers as $q$), (ii) involving only relations from $V$ and (iii) minimal, in the sense mentioned above.*

We are now ready to define our view selection problem, which relies on candidate view sets:

DEFINITION 2.3 (CANDIDATE VIEW SET). *Let $Q$ be a set of RDF queries. A candidate view set for $Q$ is a pair $\langle V, R \rangle$ such that:*

- *$V$ is a set of RDF views,*
- *$R$ is a set of rewritings such that: (i) for every query $q \in Q$ there exists exactly one rewriting $r \in R$ of $q$ using the views in $V$; (ii) all $V$ views are useful, i.e., every view $v \in V$ participates to at least one rewriting $r \in R$.*

We consider a *cost estimation function* $c^\epsilon$ which returns a quantitative measure of the costs associated to a view set. The lower the cost, the better the candidate view set is. Our cost components include the effort to evaluate the view-based query rewritings, the total space occupancy of the views and the view maintenance costs as data changes. More details about $c^\epsilon$ are provided in Section 3.3.

DEFINITION 2.4 (VIEW SELECTION PROBLEM). *Let $Q = \{q_1, q_2, \ldots, q_n\}$ be a set of RDF queries and $c^\epsilon$ be a cost estimation function. The view selection problem consists in finding a candidate view set $\langle V, R \rangle$ for $Q$ such that, for any other candidate view set $\langle V', R' \rangle$ for $Q$: $c^\epsilon(\langle V, R \rangle) \leq c^\epsilon(\langle V', R' \rangle)$.*

## 3. THE SPACE OF CANDIDATE VIEW SETS

This Section describes our approach for modeling the space of possible candidate view sets. Section 3.1 introduces the notion of a state to model one such set, while Section 3.2 presents a set of transitions that can be used to transform one state to another. Finally, Section 3.3 shows how to assign a cost estimation to each state.

### 3.1 States

We use the notion of *state* to model a candidate view set together with the rewritings of the workload queries based on these views. The set of all possible candidate view sets, then, is modeled as a set of states, which we adapt from the previous work on materialized view selection in a relational data warehouse [21]. From here forward, given a workload $Q$, we may use $S(Q)$ (possibly with subscripts or superscripts) to denote a candidate view set for $Q$. To ease the exposition, we also employ from [21] a visual representation of each state by means of a *state graph*.

DEFINITION 3.1 (STATE GRAPH). *Given a query set $Q$ and a state $S_i(Q) = \langle V_i, R_i \rangle$, the state graph $G(S_i) = (N_i, E_i)$ is a directed multigraph such that:*

- *each triple $t_i$ appearing in a view $v \in V_i$ is represented by a node $n_i \in N_i$;*
- *let $t_i$ and $t_j$ be two triples in a view $v \in V_i$, and a join on their attributes $t_i.a_i$ and $t_j.a_j$ (where $a_i, a_j \in \{s, p, o\}$). For each such join, there is an edge $e_i \in E_i$ connecting the respective nodes $n_i, n_j \in N_i$ and labeled $v{:}n_i.a_i = n_j.a_j$. We call $e_j$ a join edge;*
- *let $t_i$ be a triple in a view $v \in V_i$ and $n_i \in N_i$ be its corresponding node. For every constant $c_i$ that appears in the attribute $a_i \in \{s, p, o\}$ of $t_i$, an edge labeled $v{:}n_i.a_i = c_i$ connects $n_i$ to itself. Such an edge is called selection edge.*

The *graph of $v$* is defined as the subgraph of $G(S_i)$ corresponding to $v$. Observe that in a view, two nodes may be connected by several join edges if their corresponding atoms are connected by more than one join predicates.

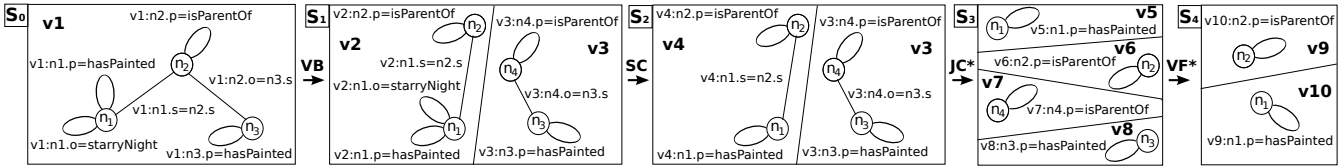

**Figure 1: Sample initial state graph $S_0$, and states attained through successive transitions.**

We define two states to be *equivalent* if they have the same view sets. Furthermore, to avoid a blow-up in the storage space required by the views, *we do not consider views including Cartesian products*. In a relational setting, some Cartesian products, e.g., between small dimension tables in an OLAP context, may not raise performance issues. In contrast, in the RDF context where all data lies in a single large triple table, views with Cartesian products are likely not interesting and their storage overhead is prohibitive. The absence of Cartesian products from our views entails that *the graph of every view is a connected component of the state graph*.

As a (simple) example, consider the state $S_0(Q) = \langle \{v_1\}, R_0 \rangle$, where $Q = \{q_1\}$ is a workload containing only the previously introduced query $q_1$, and $v_1 = q_1$. The rewriting set $R_0$ consists of the trivial rewriting $\{q_1 = v_1\}$. The graph $G(S_0)$ is depicted at left in Figure 1, and since it corresponds to a single view, it comprises only one connected component.

## 3.2 State Transitions

To enumerate candidate view sets (or, equivalently, states), we use four transitions, inspired from [21]. As we show in Section 5.1, our transition set is complete, i.e., all possible states for a given workload can be reached through our four transitions. The first three transitions remove predicates from views, thus can be seen as "relaxing", and may split a view in two, increasing the number of views. The last one factorizes two views into one, thus reducing the number of workload views. The graphs corresponding to the states before and after each transition are illustrated in Figure 1.

We use $v$:$e$ to denote an edge $e$ belonging to the view $v$ in a state graph. While we define rewritings as conjunctive queries, for ease of explanation, we now denote rewritings by (equivalent) relational algebra expressions. We use $\sigma_e$ to denote a selection on the condition attached to the edge $e$ in a view set graph. Since the query set $Q$ is unchanged across all transitions, we omit it for readability.

DEFINITION 3.2 (VIEW BREAK (VB)). *Let $S = \langle V, R \rangle$ be a state, $v$ a view in $V$ and $N_v$ the set of nodes of the graph of $v$ with $|N_v| > 2$. Let $N_{v_1}, N_{v_2}$ be two subsets of $N_v$ such that:*

- $N_{v_1} \nsubseteq N_{v_2}$ *and* $N_{v_2} \nsubseteq N_{v_1}$;
- $N_{v_1} \cup N_{v_2} = N_v$;
- *the subgraph of the graph of $v$ defined by $N_{v_1}$ (respectively, by $N_{v_2}$) and the edges between these nodes is connected.*

*We create two new views, $v_1$ and $v_2$. View $v_1$ (respectively $v_2$) derives from the graph of $v$ by copying the nodes corresponding to $N_{v_1}$ ($N_{v_2}$) and the edges between them. The head variables of $v_1$ ($v_2$) are those of $v$ appearing also in the body of $v_1$ ($v_2$), together with all additional variables appearing in the nodes $N_{v_1} \cap N_{v_2}$.*

*The new state $S' = \langle V', R' \rangle$ consists of:*

- $V' = (V \setminus \{v\}) \cup \{v_1, v_2\}$,
- $G(S')$ *is obtained from $G(S)$ by removing the graph of $v$ and adding those of $v_1$ and $v_2$, and*
- $R'$ *is obtained from $R$ by replacing all the occurrences of $v$, with $\pi_{head(v)}(v_1 \bowtie v_2)$, where $\bowtie$ is the natural join.*

For example, we apply a view break on the view $v_1$ of state $S_0$ introduced in the previous Section, and obtain the new state $S_1$:

$$S_1 = \langle \{v_2, v_3\}, \{q_1 = \pi_{head(v_1)}(v_2 \bowtie v_3)\} \rangle$$

DEFINITION 3.3 (SELECTION CUT (SC)). *Let $S = \langle V, R \rangle$ be a state and $v$:$e$ be a selection edge in $G(S)$. A selection cut on $e$ yields a state $S' = \langle V', R' \rangle$ such that:*

- $V'$ *is obtained from $V$ by replacing $v$ with a new view $v'$, in which the constant in the selection edge $e$ has been replaced with a fresh head variable (i.e., is returned by $v'$, along with the variables returned by $v$),*
- $G(S')$ *is obtained from $G(S)$ by removing the graph of $v$ and adding the one of $v'$, and*
- $R'$ *is obtained from $R$ by replacing all occurrences of $v$ with the expression $\pi_{head(v)}(\sigma_e(v'))$.*

For instance, we apply a selection cut on the edge labeled $v_2$:$n_1.o = starryNight$ of $G(S_1)$ and obtain the state $S_2$, in which $v_2$ is replaced by a new view $v_4$:

$$S_2 = \langle \{v_3, v_4\},$$
$$\{q_1 = \pi_{head(v_1)}(\pi_{head(v_2)}(\sigma_{n_1.o=starryNight}(v_4)) \bowtie v_3)\} \rangle$$

DEFINITION 3.4 (JOIN CUT (JC)). *Let $S = \langle V, R \rangle$ be a state and $v$:$e$ be a join edge in $G(S)$ of the form $n_i.c_i = n_j.c_j$, such that $c_i, c_j \in \{s, p, o\}$. A join cut on $e$ yields a state $S' = \langle V', R' \rangle$, obtained as follows:*

1. *If the graph of $v$ is still connected after the cut, $V'$ is obtained from $V$ by replacing $v$ with a new view $v'$ in which the variable corresponding to the join edge $e$ becomes a head variable, and the occurrence of that variable corresponding to $n_i.c_i$ is replaced by a new fresh head variable. The new rewriting set $R'$ is obtained from $R$ by replacing $v$ by $\pi_{head(v)}(\sigma_e(v'))$. The new graph $G(S')$ is obtained from $G(S)$ by removing the graph of $v$ and adding the one of $v'$.*

2. *If the graph of $v$ is split in two components, $V'$ is obtained from $V$ by replacing $v$ with two new symbols $v'_1$ and $v'_2$, each corresponding to one component. In each of $v'_1$ and $v'_2$, the join variable of $e$ becomes a head variable. The new rewriting set $R'$ is obtained from $R$ by replacing $v$ by $\pi_{head(v)}(v'_1 \bowtie_e v'_2)$. The new graph $G(S')$ is obtained from $G(S)$ by removing the graph of $v$ and adding the ones of $v'_1$ and $v'_2$.*

For example, cutting the join edge $v_4$:$n_1.s = n_2.s$ of $G(S_2)$ disconnects the graph of $v_4$, resulting in two new views, $v_5$ and $v_6$ (see Figure 1). View symbol $v_4$ is replaced in the rewritings by the expression $\pi_{head(v_4)}(v_5 \bowtie_{n_1.s=n_2.s} v_6)$. If we continue by cutting the edge $v_3$:$n_4.o = n_3.s$, $v_3$ is split into $v_7$ and $v_8$. The resulting state $S_3$ is:

$$S_3 = \langle \{v_5, v_6, v_7, v_8\},$$
$$\{q_1 = \pi_{head(v_1)}(\pi_{head(v_2)}(\sigma_{n_1.o=starryNight}(\pi_{head(v_4)}($$
$$v_5 \bowtie_{n_1.s=n_2.s} v_6))) \bowtie \pi_{head(v_3)}(v_7 \bowtie_{n_4.o=n_3.s} v_8)\} \rangle$$

DEFINITION 3.5 (VIEW FUSION (VF)). *Let $S = \langle V, R \rangle$ be a state and $v_1, v_2$ be two views in $V$ such that their respective graphs are isomorphic (their bodies are equivalent up to variable renaming). We denote by $\langle i \to j \rangle$ the renaming of the variables of $v_i$ into those of $v_j$. Let $v_3$ be a copy of $v_1$, such that $head(v_3) = head(v_1) \cup head(v_{2\langle 2 \to 1 \rangle})$. Fusing $v_1$ and $v_2$ leads to a new state $S' = \langle V', R' \rangle$ obtained as follows:*

- $V' = (V \setminus \{v_1, v_2\}) \cup \{v_3\}$,
- $G(S')$ *is obtained from* $G(S)$ *by removing the graphs of* $v_1$ *and* $v_2$ *and adding that of* $v_3$, *and*
- $R'$ *is obtained from* $R$ *by replacing any occurrence of* $v_1$ *with* $\pi_{head(v_1)}(v_3)$, *and of* $v_2$ *with* $\pi_{head(v_2)}(v_{3\langle 3 \to 2 \rangle})$

For example, in state $S_3$, the graphs of $v_5$ and $v_8$ are isomorphic, and can thus be fused creating the new view $v_9$. Similarly, $v_6$ and $v_7$ can be fused into a new view $v_{10}$ leading to state $S_4$.

Our transitions adapt those introduced in [21] to our RDF view selection context; the differences are detailed in [25].

### 3.3 Estimated State Cost

To each state, we associate a *cost estimation* $c^\epsilon$, taking into account: the space occupancy of all the materialized views, the cost of evaluating the workload query rewritings, and the cost associated to the maintenance of the materialized views.

For any conjunctive query or view $v$, we use $len(v)$ to denote the number of atoms in $v$, $|v|$ for the number of tuples in $v$ and $|v|^\epsilon$ for our *estimation* of this number. Let $S(Q) = \langle V, R \rangle$ be a state.

**View space occupancy (VSO$^\epsilon$)** To estimate the cardinality of a given view $v \in V$, we adopt the solution of [16], which consists in counting and storing the exact number of tuples $(i)$ for each given $s$, $p$ and $o$ value; $(ii)$ for each pair of $(s,p)$, $(s,o)$ and $(s,p)$ values. This leads to exact cardinality estimations for any 1-atom view with 1 or 2 constants. The size of an 1-atom view with no constants is the size of the data set; three-constants atoms are disallowed in our framework since they introduce Cartesian products in views.

We now turn to the case of multi-atom views. From each view $v \in V$, and each atom $t_i \in v$, $1 \leq i \leq len(v)$, let $v^i$ be the conjunctive query whose body consists of exactly the atom $t_i$ and whose head projects the variables in $t_i$. From our gathered statistics, we know $|v^i|$. We assume that values in each triple table column are *uniformly distributed*, and that values of different columns are *independently distributed*[1]. For the $s$, $p$ and $o$ columns, moreover, we store the number of distinct values, as well as the minimum and maximum values. Then, we compute $|v|^\epsilon$ based on the exact counts $|v^i|$ and the above assumptions and statistics, applying known relational formulas [18]. Finally, we use the average size of a subject, property, respectively object, the attributes in the head of $v$, and $|v|^\epsilon$, to estimate the space occupancy of view $v$.

Since the workload is known, we gather only the statistics needed for this workload: $(i)$ we count the triples matching each of the query atoms $(ii)$ we also count the triples matching all *relaxations* of these atoms, obtained by removing constants (as Sc does during the search). Consider, for instance, the following query:

$$q(X_1, X_2) \colon - t(X_1, rdf{:}type, picture), t(X_1, isLocatIn, X_2)$$

We count the triples matching the two query atoms:

$q^1(X_1) \colon - t(X_1, rdf{:}type, picture), q^2(X_1, X_2) \colon - t(X_1, isLocatIn, X_2)$ as well as the triples matching three *relaxed atoms*, obtained by removing the constants from $q^1$ and $q^2$:

$q^3(X_1, X_2) \colon - t(X_1, rdf{:}type, X_2), q^4(X_1, X_2) \colon - t(X_1, X_2, picture),$
$q^5(X_1, X_2, X_3) \colon - t(X_1, X_2, X_3)$.

Based on the cardinalities of the above atoms, we can estimate the cardinality of any possible view created throughout the search.

**Rewriting evaluation cost (REC$^\epsilon$)** This cost estimation reflects the processing effort needed to answer the workload queries using the proposed rewritings in $R$. It is computed as:

$$REC^\epsilon(S) = \sum_{r \in R} (c_1 \cdot io^\epsilon(r) + c_2 \cdot cpu^\epsilon(r))$$

---
[1] A very recent work [14] provides an RDF query size estimation method which does not make the independence assumption. This estimation method could easily be integrated in our framework.

where $io^\epsilon(r)$ and $cpu^\epsilon(r)$ estimate the I/O cost and the CPU processing cost of executing the rewriting $r$ respectively, and $c_1, c_2$ are some weights. The I/O cost estimation is:

$$io^\epsilon(r) = \sum_{v \in r} |v|^\epsilon$$

where $v \in r$ denotes a view appearing in the rewriting $r$.

The CPU cost estimation $cpu^\epsilon(r)$ sums up the estimated costs of the selections, projections, and joins required by the rewriting $r$, computed based on the view cardinality estimations and known formulas from the relational query processing literature [18].

**View maintenance cost (VMC$^\epsilon$)** The cost of maintaining the views in $V$ when the data is updated depends on the algorithm implemented to propagate the updates. In a conservative way, we chose to account only for the costs of writing/removing tuples to/from the views due to an update, ignoring the other maintenance operation costs. Consider the addition of a triple $t_+$ to the triple table, and a view $v$ of $len(v)$ atoms. With some simplification, we consider that $t_+$ joins with $f_1$ existing triples for some constant $f_1$, the tuples resulting from this, in turn, join with $f_2$ existing triples etc. Adding the triple $t_+$ thus causes the addition of $f_1 \cdot f_2 \cdot \ldots \cdot f_{len(v)}$ tuples to $v$. A similar reasoning holds for deletions. To avoid estimating $f_1, f_2, \ldots, f_{len(v)}$, which may be costly or impossible for triples which will be added in the future, we consider a single user-provided factor $f$, and compute:

$$VMC^\epsilon(S) = \sum_{v \in V} f^{len(v)}$$

The **estimated cost** $c^\epsilon$ of a state $S$ is defined as:

$$c^\epsilon(S) = c_s \cdot VSO^\epsilon(S) + c_r \cdot REC^\epsilon(S) + c_m \cdot VMC^\epsilon(S)$$

where the numerical weights $c_s$, $c_r$ and $c_m$ determine the importance of each component: if storage space is cheap $c_s$ can be set very low, if the triple table is rarely updated $c_m$ can be reduced etc.

**Impact of transitions on the cost** Transition Sc increases the view size and adds to some rewritings the CPU cost of the selection. Thus, Sc always increases the state cost. Transitions Jc and VB may increase or decrease the space occupancy, and add the costs of a join to some rewritings. Jc decreases maintenance costs, whereas VB may increase or decrease it. Overall, Jc and VB may increase or decrease the state cost. Finally, VF decreases the view space occupancy and view maintenance costs. Query processing costs may remain the same or be reduced, but they cannot increase. Thus, VF always reduces the overall cost of a state.

## 4. VIEW SELECTION & RDF REASONING

The approach described so far does not take into consideration the implicit triples that are intrinsic to RDF and that complete query answers. Section 4.1 introduces the notion of RDF entailment to which such triples are due. Section 4.2 presents the two main methods for processing RDF queries when RDF entailment is considered, namely *database saturation* and *query reformulation*. In particular, we devise a novel reformulation algorithm extending the state of the art. Finally, Section 4.3 details how we take RDF entailment into account in our view selection approach.

### 4.1 RDF entailment

The W3C RDF recommendation [26] provides a set of *entailment rules*, which lead to deriving new implicit (or *entailed*) triples from an RDF database. We provide here an overview of these rules.

Some implicit triples are obtained by generalizing existing triples using blank nodes. For instance, a triple $(s, p, o)$ entails the triple $(\_:b, p, o)$, where $s$ is a URI and $\_:b$ denotes a blank node.

Some other rules derive implicit triples from the semantics of a few special URIs, which are part of the RDF standard, and are assigned special meaning. For instance, RDF provides the rdfs:Class URI whose semantics is the set of all RDF-specific (predefined)

| Semantic relationship | RDF notation | FOL notation |
|---|---|---|
| Class inclusion | $(c_1, \text{rdfs:subClassOf}, c_2)$ | $\forall X (c_1(X) \Rightarrow c_2(X))$ |
| Property inclusion | $(p_1, \text{rdfs:subPropertyOf}, p_2)$ | $\forall X \forall Y (p_1(X,Y) \Rightarrow p_2(X,Y))$ |
| Domain typing of a property | $(p, \text{rdfs:domain}, c)$ | $\forall X \forall Y (p(X,Y) \Rightarrow c(X))$ |
| Range typing of a property | $(p, \text{rdfs:range}, c)$ | $\forall X \forall Y (p(X,Y) \Rightarrow c(Y))$ |

**Table 1: Semantic relationships expressible in an RDFS.**

*and* user-defined URIs denoting classes to which resources may belong. When, for example, a triple states that a resource $u$ belongs to a given user-defined class *painting*, i.e., $(u, \text{rdf:type}, painting)$ using the predefined URI rdf:type, an implicit triple states that *painting* is a class: $(painting, \text{rdf:type}, \text{rdfs:Class})$.

Finally, some rules derive implicit triples from the semantics encapsulated in an *RDF Schema* (RDFS for short). An RDFS specifies semantic relationships between classes and properties used in descriptions. Table 1 shows the four semantic relationships allowed in RDF, together with their first-order logic semantics. Some rules derive implicit triples through the transitivity of class and property inclusions, and of inheritance of domain and range typing. For instance, if *painting* is a subclass of *masterpiece*, i.e., $(painting, \text{rdfs:subClassOf}, masterpiece)$, which is a subclass of *work*, i.e., $(masterpiece, \text{rdfs:subClassOf}, work)$, then an entailed triple is $(painting, \text{rdfs:subClassOf}, work)$. If *hasPainted* is a subproperty of *hasCreated*, i.e., $(hasPainted, \text{rdfs:subPropertyOf}, hasCreated)$, the ranges of which are the classes *painting* and *masterpiece* respectively, i.e., $(hasPainted, \text{rdfs:range}, painting)$ and $(hasCreated, \text{rdfs:range}, masterpiece)$, then those triples are implicit: $(hasPainted, \text{rdfs:range}, masterpiece)$, $(hasPainted, \text{rdfs:range}, work)$, and $(hasCreated, \text{rdfs:range}, work)$. Some other rules use the RDFS to derive implicit triples by propagating values (URIs, blank nodes, and literals) from subclasses and subproperties to their superclasses and superproperties, and from properties to classes typing their domains and ranges. If a resource $u$ has painted something, i.e., $(u, hasPainted, \_:b)$, implicit triples are: $(u, hasCreated, \_:b)$, $(\_:b, \text{rdf:type}, painting)$, $(\_:b, \text{rdf:type}, masterpiece)$, and $(\_:b, \text{rdf:type}, work)$.

Returning complete answers requires considering all the implicit triples. In practice, RDF data management frameworks (e.g., Jena[2]) allow specifying the subset of RDF entailment rules w.r.t. which completeness is required. This is because the implicit triples brought by some rules, e.g., generalization of constants into blank nodes, may not be very informative in most settings. Of particular interest among all entailment rules are usually those derived from an RDFS, since they encode application domain semantics.

### 4.2 RDF entailment and query answering

We consider here the two main approaches previously proposed to answer queries w.r.t. a given set of RDF entailment rules: database saturation and query reformulation.

**Database saturation** The first approach *saturates* the database by adding to it all the implicit triples specified in the RDF recommendation [26]. The benefit of saturation is that standard query evaluation techniques for plain RDF can be applied on the resulting database to compute complete answers [27]. Saturation also has drawbacks. First, it needs more space to store the implicit triples, competing with the data and the materialized views. Observe that saturation adds all implicit triples to the store, whether user queries need them or not. Second, the maintenance of a saturated database,

---

[2] http://jena.sourceforge.net/

---

**Algorithm 1:** Reformulate$(q, \mathcal{S})$

**Input** : an RDF schema $\mathcal{S}$ and a conjunctive query $q$ over $\mathcal{S}$
**Output**: a union of conjunctive queries $ucq$ such that for any database $D$:
evaluate$(q, \text{saturate}(D, \mathcal{S}))$ = evaluate$(ucq, D)$

1   $ucq \leftarrow \{q\}$, $ucq' \leftarrow \emptyset$
2   **while** $ucq \neq ucq'$ **do**
3     $ucq' \leftarrow ucq$
4     **foreach** *conjunctive query* $q' \in ucq'$ **do**
5       **foreach** *atom $g$ in $q'$* **do**
6         **if** $g = t(\text{s}, \text{rdf:type}, c_2)$ *and* $c_1$ rdfs:subClassOf $c_2 \in \mathcal{S}$ **then**
7           $ucq \leftarrow ucq \cup \{q'_{[g/t(\text{s},\text{rdf:type},c_1)]}\}$ //rule 1
8         **if** $g = t(\text{s}, p_2, \text{o})$ *and* $p_1$ rdfs:subPropertyOf $p_2 \in \mathcal{S}$ **then** $ucq \leftarrow ucq \cup \{q'_{[g/t(\text{s},p_1,\text{o})]}\}$ //rule 2
9         **if** $g = t(\text{s}, \text{rdf:type}, c)$ *and* $p$ rdfs:domain $c \in \mathcal{S}$ **then**
10        $ucq \leftarrow ucq \cup \{q'_{[g/\exists X\ t(\text{s},p,X)]}\}$ //rule 3
11         **if** $g = t(\text{o}, \text{rdf:type}, c)$ *and* $p$ rdfs:range $c \in \mathcal{S}$ **then**
12        $ucq \leftarrow ucq \cup \{q'_{[g/\exists X\ t(X,p,\text{o})]}\}$ //rule 4
13         **if** $g = t(\text{s}, \text{rdf:type}, X)$ *and* $c_1, c_2, \ldots, c_n$ *are all the classes in $\mathcal{S}$* **then** //rule 5
14        $ucq \leftarrow ucq \cup \bigcup_{i=1}^{n} \{(q'_{[g/t(\text{s},\text{rdf:type},c_i)]})_{\sigma=[X/c_i]}\}$
15         **if** $g = t(\text{s}, X, \text{o})$ *and* $p_1, p_2, \ldots, p_m$ *are all the properties in $\mathcal{S}$* **then** //rule 6
16        $ucq \leftarrow ucq \cup \bigcup_{i=1}^{m} \{(q'_{[g/t(\text{s},p_i,\text{o})]})_{\sigma=[X/p_i]}\} \cup \{(q'_{[g/t(\text{s},\text{rdf:type},\text{o})]})_{\sigma=[X/rdf:type]}\}$

17 **return** $ucq$

which can be seen as an inflationary fixpoint, when adding or removing data and/or RDFS statements may be complex and costly. Finally, saturation is not always possible, e.g., when querying is performed at a client with no write access to the database.

**Query reformulation** The second approach *reformulates* a (conjunctive) query into an equivalent union of (conjunctive) queries. The complete answers of the initial query (w.r.t. the considered RDF entailment rules) can be obtained by standard query evaluation techniques for plain RDF [27] using this union of queries against the non-saturated database.

The benefit of reformulation is leaving the database unchanged. However, reformulation has an overhead at query evaluation time.

**Query reformulation w.r.t. an RDFS** Query reformulation algorithms have been investigated in the literature for the well-known *Description Logic fragment* of RDF [3, 5]: datasets with RDFSs, without blank nodes, and where RDF entailment only considers the rules associated to an RDFS (those of the third kind in Section 4.1). However, these algorithms allow reformulating queries from a strictly less expressive language than the one of our RDF queries (see Section 7 for more details) and, thus, cannot be applied to our setting. We therefore propose the Algorithm 1 that fully captures our query language, so that we can obtain the complete answers of any RDF query by evaluating its reformulation.

The algorithm uses the set of rules of Figure 2 to *unfold* the queries; in this Figure and onwards, we denote by s, p, respectively, o, a placeholder for either a constant or a variable occurring in the subject, property, respectively, object position of a triple atom. Notice that rules (1)-(4) follow from the four rules of Table 1. The evaluate and saturate functions, used in Algorithm 1 provide, respectively, the standard query evaluation for plain RDF, and the saturation of a data set w.r.t. an RDFS (Table 1). Moreover, $q_{[g/g']}$ is the result of replacing the atom $g$ of the query $q$ by the atom $g'$ and $q_{\sigma=[X/c]}$ is the result of replacing any occurrence of the variable $X$ in $q$ with the constant $c$.

$t(\text{s}, rdf\text{:}type, c_1) \Rightarrow t(\text{s}, rdf\text{:}type, c_2),$
$\qquad\qquad\qquad\text{with } c_1 \; rdfs\text{:}subClassOf \; c_2 \in \mathcal{S} \quad (1)$

$t(\text{s}, p_1, \circ) \Rightarrow t(\text{s}, p_2, \circ), \text{with } p_1 \; rdfs\text{:}subPropertyOf \; p_2 \in \mathcal{S} \quad (2)$

$t(\text{s}, p, X) \Rightarrow t(\text{s}, rdf\text{:}type, c), \text{with } p \; rdfs\text{:}domain \; c \in \mathcal{S} \quad (3)$

$t(X, p, \circ) \Rightarrow t(\circ, rdf\text{:}type, c), \text{with } p \; rdfs\text{:}range \; c \in \mathcal{S} \quad (4)$

$t(\text{s}, rdf\text{:}type, c_i) \Rightarrow t(\text{s}, rdf\text{:}type, X), \text{for any class } c_i \text{ of } \mathcal{S} \quad (5)$

$t(\text{s}, p_i, \circ) \Rightarrow t(\text{s}, X, \circ), \text{for any property } p_i \text{ of } \mathcal{S} \text{ and } rdf\text{:}type \quad (6)$

**Figure 2: Reformulation rules for an RDFS $\mathcal{S}$.**

More precisely, Algorithm 1 uses the rules in Figure 2 to generate new queries from the original one, by a backward application of the rules on the query atoms. It then applies the same procedure on the newly obtained queries until no new queries can be constructed, and then outputs the union of the generated queries. The inner loop (lines 5-16) comprises six *if* statements, one for each of the six rules above. The conditions of these statements represent the heads (right parts) of the rules, whereas the consequents correspond to their bodies (left parts). In each iteration, when a query atom matches the condition of an *if*, the respective rule is triggered, replacing the atom with the one of the body of the rule. Note that rules 5 and 6 need to bind a variable $X$ of an atom to a constant $c_i$, $p_i$, or $rdf\text{:}type$, thus use $\sigma$ to bind all the occurrences of $X$ in the query in order to retain the join on $X$ within the whole new query.

THEOREM 4.1 (TERMINATION OF **Reformulate**$(q, \mathcal{S})$). *Given a query $q$ over an RDFS $\mathcal{S}$, **Reformulate**$(q, \mathcal{S})$ terminates and outputs a union of no more than $(2|\mathcal{S}|^2)^m$ queries, where $|\mathcal{S}|$ is the number of statements in $\mathcal{S}$ and $m$ the number of atoms in $q$.*

The proof can be found in the technical report [25]. This theorem also exhibits that the query reformulation is polynomial in the size of the schema and exponential in the size of the query.

THEOREM 4.2 (CORRECTNESS OF ALGORITHM 1). *Let ucq be the output of **Reformulate**$(q, \mathcal{S})$, for a query $q$ over an RDFS $\mathcal{S}$. For any database $D$ associated to $\mathcal{S}$:*

evaluate$(q, \text{saturate}(D, \mathcal{S})) = $ evaluate$(ucq, D)$.

Again the proof is delegated to the technical report [25].

### 4.3 View selection aware of RDF entailment

We now discuss possible ways to take RDF implicit triples into account in our view selection approach. As will be explained, the exact way cardinality statistics are collected for each view atom, described first in Section 3.3, play an important role here.

**Database saturation** If the database is saturated prior to view selection, the collected statistics do reflect the implicit triples.

**Pre-reformulation** Alternatively, one could reformulate the query workload and then apply our search on the new workload. To do so, we extend the definition of our initial state, as well as our rewriting language to that of *unions* of conjunctive queries. More precisely, given a set of queries $Q = \{q_1, \ldots, q_n\}$, and assuming that Reformulate$(q_i, \mathcal{S}) = \{q_i^1, \ldots, q_i^{n_i}\}$, it is sufficient to define $S_0(Q) = \langle V_0, R_0 \rangle$ as the set of conjunctive views $V_0 = \bigcup_{i=1}^{n} \{q_i^1, \ldots, q_i^{n_i}\}$ and the set of rewritings $R_0 = \bigcup_{i=1}^{n} \{q_i = q_i^1 \cup \cdots \cup q_i^{n_i}\}$. In this case, statistics are collected on the original (non-saturated) database for the reformulated queries.

As stated in Theorem 4.1, query reformulation can yield a significant number of new queries, increasing the number of views of our initial state and leading to a serious increase of the search space. As an example, the following simple query on the Barton [24] dataset

$q(X_1, X_2, X_3)\text{:-}t(X_1, rdf\text{:}type, text), t(X_1, relatedTo, X_2),$
$\qquad\qquad t(X_2, rdf\text{:}type, subjectPart), t(X_1, language, fr),$
$\qquad\qquad t(X_2, description, X_3)$

| $q^{1,\mathcal{S}}$ | $q^1(X_1)$ | $\text{:-}t(X_1, rdf\text{:}type, picture)$ | (1) |
|---|---|---|---|
| | $\cup\, q^1(X_1)$ | $\text{:-}t(X_1, rdf\text{:}type, painting)$ | (2) |
| $q^{4,\mathcal{S}}$ | $q^4(X_1, X_2)$ | $\text{:-}(X_1, X_2, picture)$ | (1) |
| | $\cup\, q^4(X_1, isLocatIn)$ | $\text{:-}t(X_1, isLocatIn, picture)$ | (2) |
| | $\cup\, q^4(X_1, isExpIn)$ | $\text{:-}t(X_1, isExpIn, picture)$ | (3) |
| | $\cup\, q^4(X_1, rdf\text{:}type)$ | $\text{:-}t(X_1, rdf\text{:}type, picture)$ | (4) |
| | $\cup\, q^4(X_1, isLocatIn)$ | $\text{:-}t(X_1, isExpIn, picture)$ | (5) |
| | $\cup\, q^4(X_1, rdf\text{:}type)$ | $\text{:-}t(X_1, rdf\text{:}type, painting)$ | (6) |

**Table 2: Term reformulation for post-reasoning.**

is reformulated with the Barton schema into a union of 104 queries. Given the very high complexity of the exhaustive search problem (Section 5.1), such an increase may significantly impact view selection performance.

**Post-reformulation** To avoid this explosion, we propose to apply reformulation not on the initial queries but directly on the views in the final (best) state recommended by the search.

Directly doing so introduces a source of errors: since statistics are collected directly on the original database, and the queries are not reformulated, the implicit triples will not be taken into account in the cost estimation function $c^\epsilon$. To overcome this problem, we reflect implicit triples *to the statistics, by reformulating each view atom $v^i$ into a union of atoms* Reformulate$(v^i, \mathcal{S})$ *prior to the view search, and then replacing $|v^i|$ (i.e., the cardinality of $v^i$) in our cost formulas with* $|\text{Reformulate}(v^i, \mathcal{S})|$. This results in having the same statistics as if the database was saturated. Then, we perform the search using the (non-reformulated) queries and get the same best state as in the database saturation approach (as we use the same initial state and statistics). Since materializing the best state's views directly would not include the implicit triples, we need to reformulate these views first. Theorem 4.2 guarantees the correctness of post-reformulation (materializing the reformulated views on the non-saturated database is the same as materializing the non-reformulated ones on the saturated database).

Consider the query $q$ of Section 3.3, with the following schema:

$\mathcal{S} = \{painting \; rdfs\text{:}subClassOf \; picture,$
$\qquad isExpIn \; rdfs\text{:}subPropertyOf \; isLocatIn\}$

We first count (see Section 3.3) the exact number of triples matching the query atoms and their relaxed versions, namely $q^1$ to $q^5$.

We now reformulate each $q^i$ based on $\mathcal{S}$ into a union of queries, denoted $q^{i,\mathcal{S}}$. Table 2 illustrates this for $q^1$ and $q^4$ (for space reasons, we omit the other similar terms). Rule 1 (Figure 2) has been applied on $q^1$, adding to it a second union term. Applying rule 6 on $q^4$ leads to replacing $X_2$ with $isLocatIn$, $isExpIn$, and $rdf\text{:}type$ respectively in the second, third and fourth union terms of $q^{4,\mathcal{S}}$. In turn, the second term triggers rule 2 producing a fifth term, while the fourth term triggers rule 1 to produce the sixth union term.

The cardinality of each reformulated atom $q^{i,\mathcal{S}}$ is estimated prior to the search. Then, we perform the search for the non-reformulated version of $q$ using these statistics, and get the following best state:

$v_1(X_1, X_2)\text{:-}t(X_1, rdf\text{:}type, X_2), v_2(X_1, X_2)\text{:-}t(X_1, isLocatIn, X_2)$
$r_3 = \pi_{v_1.X_1, v_2.X_2}(\sigma_{X_2=picture}(v_1) \bowtie_{v_1.X_1 = v_2.X_1} v_2)$

After the search has finished, instead of the recommended views $v_1$ and $v_2$, we materialize their reformulated variants $v_1'$ and $v_2'$:

$v_1'(X_1, X_2)\text{:-}t(X_1, rdf\text{:}type, X_2)$
$\cup\, v_1'(X_1, painting)\text{:-}t(X_1, rdf\text{:}type, painting)$
$\cup\, v_1'(X_1, picture)\text{:-}t(X_1, rdf\text{:}type, picture)$
$\cup\, v_1'(X_1, picture)\text{:-}t(X_1, rdf\text{:}type, painting)$

$v_2'(X_1, X_2)\text{:-}t(X_1, isLocatIn, X_2)$
$\cup\, v_2'(X_1, X_2)\text{:-}t(X_1, isExpIn, X_2)$

Executing $r_3$ on $v_1'$ and $v_2'$ provides the complete answers for $q$.

In post-reformulation, finding the best state does not require saturating the database nor multiplying the queries and making the search space size explode (as pre-reformulation does). Thus, this is the best approach for situations where database saturation is not an option, which is also shown through our experiments in Section 6.5.

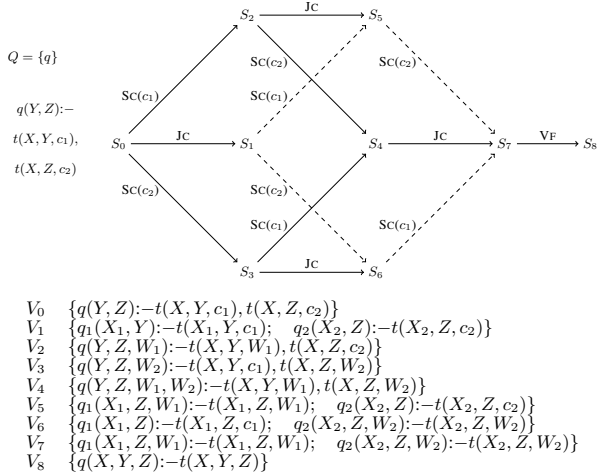

| | |
|---|---|
| $V_0$ | $\{q(Y,Z):-t(X,Y,c_1),t(X,Z,c_2)\}$ |
| $V_1$ | $\{q_1(X_1,Y):-t(X_1,Y,c_1); \quad q_2(X_2,Z):-t(X_2,Z,c_2)\}$ |
| $V_2$ | $\{q(Y,Z,W_1):-t(X,Y,W_1),t(X,Z,c_2)\}$ |
| $V_3$ | $\{q(Y,Z,W_2):-t(X,Y,c_1),t(X,Z,W_2)\}$ |
| $V_4$ | $\{q(Y,Z,W_1,W_2):-t(X,Y,W_1),t(X,Z,W_2)\}$ |
| $V_5$ | $\{q_1(X_1,Z,W_1):-t(X_1,Z,W_1); \quad q_2(X_2,Z):-t(X_2,Z,c_2)\}$ |
| $V_6$ | $\{q_1(X_1,Z):-t(X_1,Z,c_1); \quad q_2(X_2,Z,W_2):-t(X_2,Z,W_2)\}$ |
| $V_7$ | $\{q_1(X_1,Z,W_1):-t(X_1,Z,W_1); \quad q_2(X_2,Z,W_2):-t(X_2,Z,W_2)\}$ |
| $V_8$ | $\{q(X,Y,Z):-t(X,Y,Z)\}$ |

**Figure 3: Sample exhaustive strategy (solid arrows), EXNAÏVE strategy (solid and dashed arrows), and view sets corresponding to each state.**

## 5. SEARCHING FOR VIEW SETS

This Section discusses strategies for navigating in the space of candidate view sets (or states), looking for a low- or minimal-cost state. We discuss the exhaustive search strategies and identify an interesting subset of *stratified* strategies in Section 5.1, based on which we analyze the size of the search space. In Section 5.2, we present several efficient optimizations and search heuristics.

### 5.1 Exhaustive Search Strategies

We define the *initial state* of the search as $S_0(Q) = \langle V_0, R_0 \rangle$, such that $V_0 = Q$, i.e., the set of views is exactly the set of queries, and each rewriting in $R_0$ is a view scan. The state graph $G(S_0)$ corresponds to the queries in $Q$. Clearly, the rewriting cost of $S_0$ is low, since each query rewriting is simply a view scan. However, its space consumption and/or view maintenance costs may be high.

We denote by $S \xrightarrow{\tau} S'$ the application of the transition $\tau \in \{\text{SC,JC,VB,VF}\}$ on a state $S$, leading to the state $S'$.

DEFINITION 5.1 (PATH). *A path is a sequence of transitions of the form:* $S_0 \xrightarrow{\tau_0} S_1, S_1 \xrightarrow{\tau_1} S_2, \ldots, S_{k-1} \xrightarrow{\tau_{k-1}} S_k$.

For instance, in Figure 3, $(S_0 \xrightarrow{\text{SC}(c_2)} S_3), (S_3 \xrightarrow{\text{JC}} S_6)$ is a path. We may denote a path simply by its transitions, e.g., (SC($c_2$), JC).

THEOREM 5.1 (COMPLETENESS OF THE TRANSITION SET). *Given a workload $Q$ and an initial state $S_0$, for every possible state $S(Q)$, there exists a path from the initial state $S_0$ to $S$.*

The proof is given in our technical report [25].

DEFINITION 5.2 (STRATEGY). *A search strategy $\Sigma$ is a sequence of transitions of the form:*
$\Sigma = (S_{i_1} \xrightarrow{\tau_{i_1}} S'_{i_1}), (S_{i_2} \xrightarrow{\tau_{i_2}} S'_{i_2}), \ldots, (S_{i_{k-1}} \xrightarrow{\tau_{i_{k-1}}} S'_{i_{k-1}}),$
$(S_{i_k} \xrightarrow{\tau_{i_k}} S'_{i_k})$
*where $S_{i_1} = S_0$, for every $j \in [1..k]$ $\tau_{i_j} \in \{\text{SC,JC,VB,VF}\}$, and for every $j \in [2..k]$ there exists $l < j$ such that $S'_{i_l} = S_{i_j}$ (each state but $S_0$ must be attained before it is transformed).*

For example, for the one-query workload depicted at the top left of Figure 3, one possible strategy is:

**Algorithm 2:** EXNAÏVE($S_0$)

**Input**: an initial state $S_0$
**Output**: the best state $S_b$ found

1  $S_b \leftarrow S_0, S_{new} \leftarrow null, CS \leftarrow \{S_0\}, ES \leftarrow \emptyset, NS \leftarrow \emptyset$
2  **while** $CS \neq \emptyset$ **do**
3    **foreach** *state* $S_c \in CS$ **do**
4      $S_{new} \leftarrow applyTrans(\{\text{SC,JC,VB,VF}\}, S_c, (ES \cup CS))$
5      **if** $S_{new} = null$ **then** move $S_c$ from $CS$ to $ES$
6      **else**
7        $CS \leftarrow CS \cup \{S_{new}\}$
8        **if** $c^\epsilon(S_{new}) < c^\epsilon(S_b)$ **then** $S_b \leftarrow S_{new}$

$\Sigma_1 = (S_0 \xrightarrow{\text{SC}(c_1)} S_2), (S_2 \xrightarrow{\text{SC}(c_2)} S_4), (S_0 \xrightarrow{\text{SC}(c_2)} S_3),$
$(S_3 \xrightarrow{\text{SC}(c_1)} S_4), (S_0 \xrightarrow{\text{JC}} S_1)$

A strategy $\Sigma$ is *exhaustive* if any state $S$ that can be reached through a path, is also reached in $\Sigma$ (not necessarily through the same path). For instance, in Figure 3, the solid arrows depict an exhaustive strategy, reaching all possible states.

We first consider a simple family of strategies called EXNAÏVE and described through Algorithm 2. EXNAÏVE strategy (as all strategies presented in this work) maintains a *candidate state set* $CS$ and a set of *explored states* $ES$. $CS$ keeps the states on which more transitions can be possibly applied and is initially $\{S_0\}$. $ES$ is disjoint from $CS$ and is empty in the beginning. A state $S$ is explored, when any state $S' = \tau(S)$ obtained by applying some transition $\tau \in \{\text{SC, JC, VB, VF}\}$ to $S$, already belongs either to $CS$ or to $ES$. EXNAÏVE at each point picks a state $S_c$ from $CS$ and tries to apply a transition to it ($applyTrans$, line 4). If no new state is obtained, $S_c$ was already explored and is moved to $ES$ (line 5); otherwise, the newly obtained state ($S_{new}$) is copied to $CS$ (line 7). During the search, we also keep the *best state* found so far (denoted $S_b$), i.e., having the lowest cost $c^\epsilon(S)$ (line 8). The strategy stops when no new states can be found. Clearly, EXNAÏVE strategies are exhaustive. In Figure 3, the solid and dashed arrows, together, illustrate an EXNAÏVE strategy.

For a given strategy $\Sigma$, the *paths to a state* $S \in \Sigma$, denoted $\hookrightarrow S$, is the set of all $\Sigma$ paths whose final state is $S$. In an EXNAÏVE strategy there may be multiple paths to some states, e.g., $S_6$ is reached twice in our example, which slows down the search. We define the notion of *stratification* to reduce the number of such duplicate states.

DEFINITION 5.3 (STRATIFIED PATH). *A path $p \in \hookrightarrow S$ for some state $S \in \Sigma$ is stratified iff it belongs to the regular language:* VB* SC* JC* VF*.

A stratified path constrains the order among the types of transitions on the path: all possible view breaks appear only in the beginning of the path and are followed by the selection cuts. Join cuts appear only after all selection cuts are applied and are in turn followed by zero or more view fusions. In Figure 3, all solid-arrow paths starting from $S_0$ are stratified.

The following theorem formalizes the interest of stratified paths.

THEOREM 5.2 (COMPLETENESS OF STRATIFIED PATHS). *Let $Q$ be a query workload and $S(Q)$ be a state for $Q$. There exists a stratified path leading from the initial state $S_0$ to $S$.*

The proof can be found in our technical report [25].
We can now identify an interesting family of strategies.

DEFINITION 5.4 (STRATIFIED STRATEGY). *A strategy $\Sigma$ is stratified iff for any $S \in \Sigma$ and $p \in \hookrightarrow S$, $p$ is stratified.*

In Figure 3, any topological sort of the solid edges is a stratified strategy, more efficient than the EXNAÏVE one illustrated in the Figure, since the latter performs four extra transitions. Observe that a stratified strategy does not constrain the order of transitions that *are not on the same path*. For instance, in Figure 3, a stratified strategy may apply the transition $S_0 \xrightarrow{\text{JC}} S_1$ before all the SCs.

We now define the important family of EXSTR strategies. Starting from the initial state $S_0$, an EXSTR strategy picks any state on which it applies any applicable transition, *preserving the stratification of all strategy paths*. Several EXSTR strategies may exist for a workload, differing in their ordering of the transitions. We will simply use EXSTR to refer to any of them. The EXNAÏVE strategy (Algorithm 2) can be turned to an EXSTR one through the following modification: when $applyTrans$ (line 4) is called on a state $S_c$, it should apply the transitions in a stratified way, i.e., first it attempts a VB and only if no new state is obtained, it applies an SC, and then a JC and, finally, a VF.

THEOREM 5.3 (INTEREST OF EXSTR). *(i) Any* EXSTR *strategy is exhaustive. (ii) For a given workload Q, and arbitrary* EXSTR *strategy* $\Sigma_S$ *and* EXNAÏVE *strategy* $\Sigma_N$, $\Sigma_S$ *has at most the number of transitions of* $\Sigma_N$.

The proof is given in [25]. Due to Theorem 5.3, among the exhaustive strategies, we will only consider *wlog* the stratified ones.

**Size of the search space** Let the workload $Q$ having in total $n$ nodes in its initial state $S_0$. Denoting by $B_k$ the $k$-th Bell number (the number of partitions of a set of size $k$), and by $\mu(n,k)$ the number of minimal covers with $k$ members of a set of size $n$, for the number of candidate view sets we have $NS(Q,n) \leq \sum_{k=1}^{n} 2^{kn^2} \mu(n,k) B_k$. The details are provided in [25].

**Time complexity** The time complexity of exhaustive search can be derived from the number of states created by each transition and the time complexity of the transition. The cost of a SC, JC and VB is linear in the size of the largest view, which is bound by $3n$, whereas VF requires checking query equivalence, which is in $O(2^n)$ [7].

The complexity of exhaustive search is very high and, even if views are selected off-line and thus time is not a concern, it brings real issues due to memory limitations. This highlights the need for robust strategies with low memory needs, and efficient heuristics.

## 5.2 Optimizations and heuristics

We now discuss a set of search strategies with interesting properties, as well as a set of pruning heuristics which may be used to trade off completeness for efficiency of the search.

**Depth-first search strategies** (DFS) A (stratified) strategy $\Sigma$ is depth-first iff the order of $\Sigma$'s transitions satisfies the following constraint. Let $S$ be a state reached by a path $p$ of the form VB*. Immediately after $S$ is reached, $\Sigma$ enumerates all states recursively attainable from $S$ by SC only. This process is then repeated with JC and then with VF. The pseudocode of DFS can be obtained by replacing lines 3-4 of Algorithm 2 with the following ones, where $recApplyTrans$ returns all states that can be reached by a specific transition starting from a given state:

---
**foreach** *state* $S_{\text{VB}} \in \{recApplyTrans(\text{VB}, S_0)\}$ **do**
  **foreach** *state* $S_{\text{SC}} \in \{recApplyTrans(\text{SC}, S_{\text{VB}})\}$ **do**
    **foreach** *state* $S_{\text{JC}} \in \{recApplyTrans(\text{JC}, S_{\text{SC}})\}$ **do**
      **foreach** *state* $S_{\text{VF}} \in \{recApplyTrans(\text{VF}, S_{\text{JC}})\}$ **do**
        ⋯

---

For instance, in Figure 3, the following strategy $\Sigma_3$ is DFS:
$$\Sigma_3 = (S_0 \xrightarrow{\text{SC}(c_1)} S_2), (S_2 \xrightarrow{\text{SC}(c_2)} S_4), (S_4 \xrightarrow{\text{JC}} S_7),$$
$$(S_7 \xrightarrow{\text{VF}} S_8), (S_0 \xrightarrow{\text{SC}(c_2)} S_3), (S_3 \xrightarrow{\text{JC}} S_6)$$

An advantage of DFS strategies is that they fully explore each obtained state more quickly, reducing the number of states stored in $CS$. This results in a significant reduction of the maximum memory needs during the search compared, e.g., with EXNAÏVE, which develops a huge number of candidates before fully exploring them.

**Aggressive view fusion** (AVF) This technique can be included in any strategy and is based on the fact that VF can only decrease the overall cost of a state (Section 3.3). Once a new state $S$ is obtained through some SC, JC or VB, we recursively apply on $S$ all possible VFs (until no more views can be fused). It can be shown that such repeated VFs converge to a single state $S^{\text{VF}}$. We then discard all intermediate states leading from $S$ to $S^{\text{VF}}$ and add only $S^{\text{VF}}$ to $CS$. Thus, AVF preserves the optimality of the search, all the while eliminating many intermediary states whose estimated cost is guaranteed to be higher than that of $S^{\text{VF}}$. For example, assume we reach a state $S$ containing three identical views. We apply a VF on $S$ fusing two of the three views and obtain the state $S'$. We then apply a VF on $S'$ fusing the two remaining identical views and obtain $S^{\text{VF}}$. AVF discards $S'$ and keeps only $S^{\text{VF}}$ to continue the search.

**Greedy stratified** (GSTR) This strategy starts by applying all possible VB transition sequences on $S_0$. It then discards *all the obtained states but* $S_b$, and repeatedly applies on it all possible SC. Keeping only $S_b$, it proceeds in the same way by applying JC and then VF. The interest of GSTR lies in the possibility to combine it with the AVF technique, leading to the GSTR-AVF strategy. GSTR-AVF has low memory needs due to the many states dropped by GSTR and AVF and moves fast towards lower-cost states due to AVF. Although neither GSTR nor GSTR-AVF can guarantee optimality, they perform well in practice, as our experiments show.

**Stop conditions** We use some *stop conditions* to limit the search by considering that some states are not promising and should not be explored. Clearly, stop conditions lead to non-exhaustive search. We have considered the following stop conditions for a state $S$.

- $stop_{tt}(S)$: true if a view in $S$ is the full triple table $t$.
- $stop_{var}(S)$: true if a view in $S$ has only variables. The idea is that we reject $S$ since we consider its space occupancy to be too high. This condition is not applicable if it is satisfied by the initial state, but such queries are of limited interest.
- $stop_{time}(S)$: true if the search has lasted more than a given amount of time. Observe that our approach is guaranteed to have *some* recommended $S_b$ state at any time.

## 6. EXPERIMENTAL EVALUATION

This Section presents an experimental evaluation of our approach, which we have fully implemented as a Java 6 application. The application takes as input a set of conjunctive RDF queries and possibly an RDF schema, and produces as output the set of recommended views and query rewritings. It uses a database back-end to store both the original RDF data and schema, and the views.

**Platform and data layout** We used PostgreSQL (version 8.4.3) as the database back-end for its reputation as a (free) efficient platform that has been used in several related works [1, 15, 16, 20, 23]. Integrating our view selection approach with another platform is easy as soon as that platform supports the evaluation of our select-project-join rewritings, and provided that the cost function is appropriately customized to account for the respective evaluation engine.

As in many previous works, for efficiency, we stored the data in a dictionary-encoded triple table, using a distinct integer for each distinct URI or literal appearing in an $s$, $p$ or $o$ value. The encoding dictionary was stored as a separate table indexed both by the integer dictionary code and by the encoded constant. The triple table was clustered by the columns $p$ and then $s$, to enhance the efficiency of (frequent) queries where the $p$ values are specified in most or all atoms. Moreover, we indexed the encoded triple table on $s$, $p$, $o$, and all two- and three-column combinations.

**Data and queries** As in previous works [1, 15, 23], we used the Barton RDF dataset and RDFS [24]. The initial dataset consists of about 50 million triples. After some cleaning (removing formatting errors, eliminating duplicates etc.) we kept about 35 million distinct triples. The space occupied by the encoded triple table, the dictionary and the indexes within PostgreSQL was 39 GB.

The Barton query workload [24] contains few queries with no commonality among them. To better test our approach, we built two query generators, producing queries of controllable size, shape, and commonality. The first one simply outputs the desired queries, and has maximum flexibility. The second takes as input not only the workload characteristics, but also a dataset (RDF + RDFS) and generates queries having non-empty answers on the given dataset. We used it to obtain interesting workloads on the Barton dataset.

**Weights of cost components** For $VSO$ and $REC$ (Section 3.3), we used $c_s$=1 and $c_r$=1. For each workload, we set the value of $c_m$ taking into account the database size and the average number of atoms in each query, so that for the initial state $S_0$, $c_m \cdot VMC$ is within at most two orders of magnitude from the other two cost components, $c_s \cdot VSO$ and $c_r \cdot REC$. In most cases, this lead to $c_m$=0.5. Finally, we set $f$=2 in $VMC$, since this value gave the most appropriate range to $VMC$ through the search.

**Hardware and memory** The PostgreSQL server ran on a separate 2.13 GHz Intel Xeon machine with 8GB RAM. We ran search algorithms on two classes of hardware: a *desktop* 8-core Intel Xeon 2.13 GHz machine with 16 GB RAM (the JVM was given 4 GB), and several *cluster machines*, each of which is a 4-core Intel Xeon 2.33 GHz with 4 GB RAM (the JVM was given 3 GB). Each experiment ran on one machine. While there are opportunities for parallelization (see Section 8), we did not exploit them in this work. All machines were running Mandriva Linux 2.6.31.

## 6.1 Competitor search strategies

We have implemented the three strategies, *Pruning*, *Greedy* and *Heuristic*, introduced in the relational view selection work which inspired our states and transitions [21]. All these strategies follow a divide-and-conquer approach. They start by breaking down the initial state into a set of 1-query states, and apply all possible edge removals, then all possible view breaks on each such state. Then, they seek to put back together states corresponding to the complete workload by adding up and, when appropriate, fusing, one state for each workload query. Since any combination of partial states leads to a valid state in [21], the number of states thus created explodes. To avoid it, *Pruning* discards partial states outgrowing the given space or cost budget, whereas *Greedy* develops very few states: it only keeps the best combined state, say, for the workload queries $\{q_1, q_2\}$, even though this may prevent finding the best combined state for $\{q_1, q_2, q_3\}$. Finally, *Heuristic* resembles *Pruning*, except that after having built all one-query states, it only keeps: the minimal-cost state for each query, and any states which offer some view fusing opportunity. Since our algorithms do not use a cost or space budget, we did not give one to the [21] strategies either. This does not prevent their pruning which is mostly based on comparing two states and discarding the less interesting one.

**Search strategy acronyms** In the sequel, for convenience, we will refer to the [21] strategies simply as *Pruning*, *Greedy* and *Heuristic*. Among the strategies we propose (see Section 5.2), DFS is the (stratified) depth-first search, while GSTR is the greedy strategy. The suffix -AVF after a strategy name denotes aggressive view fusion is applied in conjunction with that strategy, while the -STV suffix denotes that the $stop_{var}$ stop condition is used.

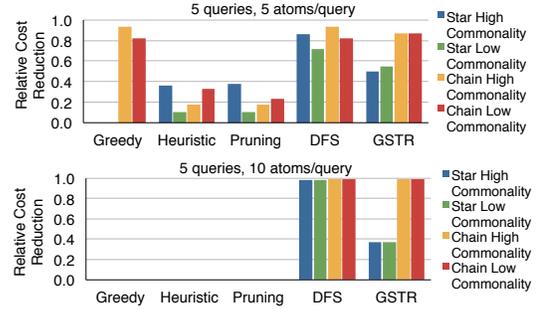

Figure 4: Strategy comparison on small workloads.

**Relative cost reduction** To assess search effectiveness, we define the *relative cost reduction (rcr)* of a given strategy $\Sigma$ and workload $Q$, at a given moment, as the ratio $(c^\epsilon(S_0) - c^\epsilon(S_b))/c^\epsilon(S_0)$, that is, the fraction of the cost of the initial state $S_0$, avoided by the current best state found by $\Sigma$ by that moment during the search.

## 6.2 Comparison with existing strategies

We compare our strategies with those of [21] for two small workloads of 5 queries each. While the queries they tested involve on average 4 relations, one needs more RDF atoms than relations to express the same logical query, since data that would fit in a wide relational tuple is split over many RDF triples. Thus, queries in the first and second workload have 5 and 10 atoms each, respectively.

Figure 4 shows the $rcr$ of the three strategies of [21] and our strategies DFS-AVF-STV and GSTR-AVF-STV. The reasons for using the specific heuristics on our strategies are explained in Section 6.3. The Figure considers workloads of star and chain queries, which are typical in RDF. In particular, star queries translate to query graphs (Definition 3.1) that are cliques (each atom is connected to all others), allowing for many VBs and JCs and, therefore, have a search space of increased size, whereas chain queries can be considered an average case regarding the difficulty of the search. The workloads were generated both with high and low commonality across queries and we used the $stop_{time}$ stop condition set to 30 minutes. While this may seem long, recall that the complexity of search is high (Section 5.1). We consider this duration acceptable as view selection is an *off-line* process. The overhead is worth it especially for large workloads, and/or queries asked repeatedly.

As can be seen in Figure 4, for the smaller workload, all strategies ran well, with DFS-AVF-STV and GSTR-AVF-STV being the best. The runs did not finish, i.e., the strategies might have found better solutions by searching longer. *Greedy* managed to reduce the cost significantly for chains queries but failed to find any state better than the initial one for stars queries. For the larger workload, the [21] strategies failed to produce any solution, as they outgrow the available memory building *partial* states (for $1, 2, 3$ queries etc.) before building *any* state covering all 5 queries. In contrast, DFS-AVF-STV and GSTR-AVF-STV keep running and achieve interesting cost reductions. The same trend was observed on workloads with cycle- and random graph-shaped queries (we generated both sparse and dense graphs), at high and low commonality.

Thus, from now on, and in particular for large workloads, we focus only on our strategies, since those of [21] systematically outgrow the memory before reaching a full candidate view set.

## 6.3 Impact of heuristics and optimizations

We now study the impact of the AVF and STV techniques on the search space explored by our algorithms. A tiny workload of 2 queries of 4 atoms each (satisfiable on the Barton dataset) suffices to illustrate this (Figure 5). The queries used are star-shaped with low commonality; [25] shows similar results for other workloads.

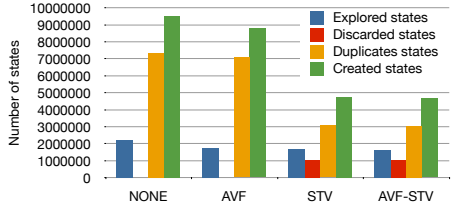

Figure 5: Impact of heuristics on the search.

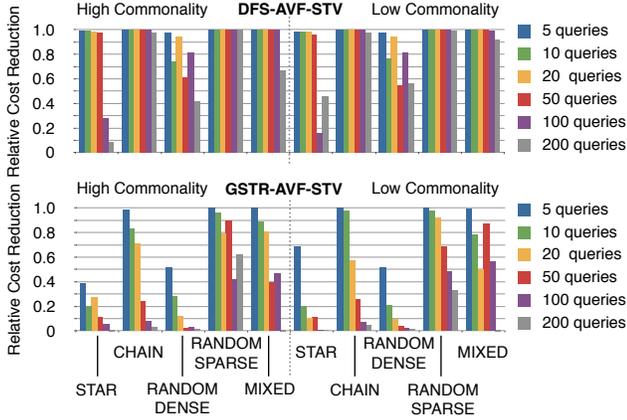

Figure 6: Relative cost reduction for large workloads.

We used the DFS strategy and several combinations of heuristics. The states *created* are those reached by the search, while *duplicates* are those already attained through a different search path (already belong to $CS$ or $ES$; see Section 5) and are ignored. *Discarded* are the states excluded from the search once they are created, whereas *explored* are the ones from which all outgoing transitions respecting DFS have been explored. In this experiment that ran in the cluster, all strategies completed execution and reached the same best state.

A first remark based on Figure 5 is that the number of duplicate states may be quite important. Duplicates occur because even when using a stratified strategy, a state may be reached by more than one path. For instance, assume for some given views $v_1, v_2$ that an SC modifies $v_1$ into $v'_1$ (denoted $v_1 \xrightarrow{\text{Sc}(c_1)} v'_1$) and similarly $v_2 \xrightarrow{\text{Sc}(c_2)} v'_2$. From the state $(v_1, v_2)$, our algorithms reach the state $(v'_1, v'_2)$ twice: once through $(v_1, v'_2)$ and a second time through $(v'_1, v_2)$. Our algorithm identifies such states as soon as they are created, in order not to repeat their exploration.

Second, Figure 5 shows that AVF (which fuses *views within one candidate set* as soon as possible) reduces the number of created states (while preserving optimality as explained in Section 5.2), because no state containing identical views is explored. A third remark is that STV discards a significant number of states, which trims down significantly all state counts. The combination AVF-STV is marginally better than STV and was efficient in all the experiments we ran. Hence, we systematically use it in the sequel.

### 6.4 Cost reduction on large workloads

We study the scalability of our DFS and GSTR algorithms for large query workloads. To this purpose, we generated workloads of 5, 10, 20, 50, 100 and 200 queries; each query has 10 atoms, i.e., the views of the initial states contains 10 atoms on average. We consider workloads consisting of: star queries only; chain queries only; random-graph shaped queries (with two variants, dense graph and sparse graph); mixed, combining queries of all the previous shapes. For each kind of workload, we generate three low- and three high-commonality variants. On each of these 30 workloads, we ran DFS-AVF-STV and GSTR-AVF-STV. We used the $stop_{time}$

| Workload $Q$ | $\lvert Q\rvert$ | $\#a(Q)$ | $\#c(Q)$ | $\lvert Q^r\rvert$ | $\#a(Q^r)$ | $\#c(Q^r)$ |
|---|---|---|---|---|---|---|
| $Q_1$ | 5 | 33 | 35 | 20 | 143 | 157 |
| $Q_2$ | 10 | 76 | 77 | 231 | 1436 | 1651 |

Table 3: Workloads used for reformulation experiments.

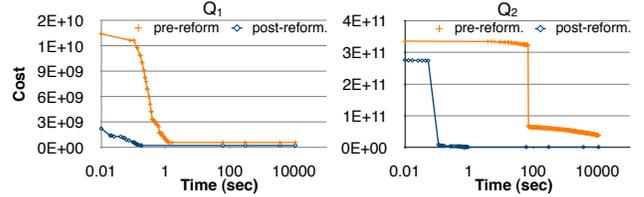

Figure 7: Search for view sets using reformulation.

stop condition set to 3 hours. These experiments ran in the cluster.

Figure 6 plots for each of the 10 workload types, the $rcr$ averaged over the 3 workloads of that type, at the end of the search. A first remark is that DFS's relative cost reduction is very impressive overall, and in many cases around 0.99. Second, note that the $rcr$ of GSTR-AVF-STV is generally smaller than that of DFS-AVF-STV, because GSTR explores significantly fewer states than DFS and might miss interesting opportunities. Third, we can distinguish "easier" workloads, such as chains and random-sparse graphs, resulting in query graphs with fewer edges and, thus, fewer transitions. For such workloads, the $rcr$ is higher since the search space is smaller (and bigger part of it was explored). Stars and random-dense graphs are difficult cases, as they lead to many edges, thus smaller $rcr$s. Finally, the $rcr$s obtained for high-commonality workloads are generally higher than for low-commonality, e.g., for random-dense and mixed workloads. This confirms the intuition that more factorization opportunities lead to higher gains. DFS-AVF-STV resulted in views with 3.2 atoms in average, whereas GSTR-AVF-STV produced views with 6.5 atoms in average.

We conclude that DFS-AVF-STV scales well up to 200 queries, depending on the workload structural complexity, and can achieve very significant reductions in the state cost.

### 6.5 View selection and implicit triples

We study the impact of implicit triples on view selection performance. Starting from a non-saturated database $D$ and workload $Q$, three scenarios are possible: $(i)$ saturated database $D^s$, search on $Q$ and the statistics of $D^s$; $(ii)$ original database $D$, search on the pre-reformulated workload $Q^r$ and the statistics of $D$; $(iii)$ original database $D$, search on $Q$ with the statistics of the saturated database $D^s$ (recall from Section 4.3 that we gather them *without* actually saturating the database). Of course, we consider the same RDF entailment rules for the three scenarios, i.e., those brought by an RDFS. Saturation and post-reformulation coincide for any search algorithm, since they lead to the same input statistics and workload. Hence, we only study the search for pre- and post-reformulation.

This experiment uses the Barton dataset as well. The schema consists of 39 classes, 61 properties, and 106 RDFS statements of the kinds listed in Table 1. We generated two satisfiable workloads $Q_1$ and $Q_2$, whose properties and those of their reformulated versions $Q_1^r$ and $Q_2^r$ are characterized in Table 3; $\lvert Q\rvert$ denotes the number of queries in $Q$, $\#a(Q)$ the number of atoms and $\#c(Q)$ the number of constants. $Q_1$ is a subset of $Q_2$.

Figure 7 shows the evolution of the best cost found by DFS-AVF-STV for both workloads (post-reformulation) and their reformulated variants (pre-reformulation). The search was cut after 3 hours. We see that the initial state for reformulated workloads has higher cost than the original workloads. Further, the best state cost decreases rapidly with post-reformulation, because the workload is much smaller and the search space is traversed faster. In contrast, the important workload sizes slow down the cost decrease for pre-reformulation. The best cost of pre-reformulation is higher than that of post-reformulation, by a factor of 2.7 for $Q_1$, and 22 for $Q_2$. This confirms our expectation that the advantages of post-reformulation are most visible for larger workloads (with larger $Q^r$). Moreover, the best cost is reached faster in post-reformulation.

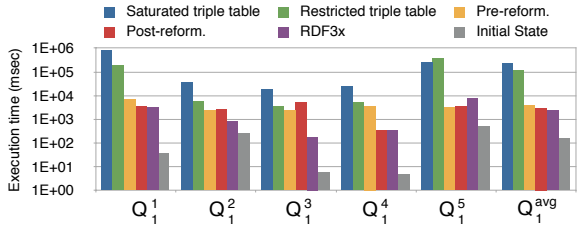

**Figure 8: Execution times for queries with RDFS.**

In general, the number of implicit triples increases with the size of the database $D$ and of the schema $S$. We show in [25] that the bound is $O(|D| \times |S|)$ for the considered RDFS entailment rules. Similarly, $|Q^r|$ may be the same as $|Q|$, or exponentially larger (Theorem 4.1). *In a reformulation-based setting*, view selection based on post-reformulation is clearly better than based on pre-reformulation, since the initial state is better and search is faster, especially for large workloads. *Among saturation and post-reformulation*, the best choice strongly depends on the context (distribution, rights to update the database, frequency and types of updates etc.) as explained in Section 4.2. The views recommended in a saturation and a post-reformulation context are the same.

### 6.6 View-based query evaluation

We now study the benefits that our recommended views actually bring to query evaluation (recall though that our view selection does not optimize for query evaluation *only*, but for a combination including storage and maintenance costs). For the workload $Q_1$ described in Section 6.5, we materialized the views recommended by pre- and post-reformulation, and ran the 5 queries $Q_1^1$ to $Q_1^5$ of $Q_1$ using $(i)$ the views, $(ii)$ the (dictionary-encoded, heavily indexed) *saturated* triple table in PostgreSQL, $(iii)$ a restricted version of $(ii)$ only with the triples needed for answering $Q_1$, $(iv)$ RDF-3X [17] (loading the saturated database in it), and $(v)$ the materialization of the query workload (initial state). RDF-3X times were put as a reference; by using PostgreSQL (even with views) we did not expect to get better times than those of the state-of-the-art RDF platform.

The views were materialized in 81 seconds for post-reformulation (the total view size was 433 MB or 15% of the database size), and 103 seconds for pre-reformulation (601 MB or 21% of the database size). Figure 8 shows that using our views, queries are evaluated more than an order of magnitude faster than on the triple table, even when using the restricted triple table $(iii)$. Both pre- and post-reformulation performed in the range of RDF-3X. This is a promising result, since our approach can be used on top of RDF-3X and achieve an even bigger gain. Finally, as expected, materializing the queries gives the best results (simply scanning the views is sufficient). More experiments are provided in [25].

Pre-computed views are likely to speed up query evaluation in any platform, simply by avoiding computations at runtime. Moreover, our framework $(i)$ avoids the overhead of query rewriting at run-time, as query rewritings are also pre-computed and $(ii)$ could easily translate our rewritings directly to any RDF platform's logical plans, exploiting its physical optimization capabilities.

### 6.7 Experiment conclusion

Our experiments have shown that the GSTR and DFS strategies scale well on up to 200 queries and achieve impressive cost reduction factors in many cases close to 99%. The strategies of [21] are also effective for small workloads, but they outgrow the memory on larger ones before producing a solution. The AVF and STV heuristics are efficient and effective, i.e., they reduce the search space while preserving view set quality. Post-reformulation largely outperforms pre-reformulation in terms of speed and effectiveness of the candidate view set selection. Finally, our recommended views do reduce query evaluation times by several orders of magnitude.

A tighter integration of the view selection tool with the internals of the data management platform, and/or using a dedicated RDF system, is likely to increase performance gains even more.

## 7. RELATED WORKS

Our work is among the first to explore materialized view selection in RDF databases. The closest works related to ours are [6] and [9]. RDFMatView [6] recommends RDF indices to materialize for a given workload, while in [9] a set of path expressions appearing in the given workload is selected to be materialized, both aiming at improving the performance of query evaluation. Unlike our approach, none of these works aims at rewriting the queries completely using the materialized indices or paths and, thus, cannot be used in scenarios where the client needs to process her queries even without access to the database. Moreover, they do not consider the implicit triples that are inherent to RDF.

Commonly used RDF management platforms (e.g., Sesame, 3store or Jena) are based on a relatively simple mapping of triples within a relational database. Many works have addressed the efficient processing of RDF queries and updates [1, 15, 16, 17, 20, 22, 23], proposing various storage and indexing models. In vertical partitioning [1] one $(s, o)$ relation is created for each property value (possibly leading to large unions for queries with variables in the $p$ position). The authors of [16, 17] have built RDF-3X, a native RDF query engine. In many of the approaches, the $(s, p, o)$ table is indexed in multiple ways (by each attribute, each pair of attributes etc.), a technique originally introduced in [23]. Recently, the problem of view-based SPARQL query rewriting was studied in [13]. These techniques have been shown to result in good RDF query and update performance. We view our approach as complementary to these works, since we seek to identify materialized views to store *on top (independently) of* the base store and indexes. To adapt our approach to a specific RDF platform, one only needs $(i)$ an execution framework capable of evaluating our simple select-project-join rewritings and $(ii)$ possibly, tailoring the cost function to the particularities of the platform. Our approach improves performance by *exploiting pre-computed results and thus avoiding computations at query evaluation time*, gains likely to extend to any context.

The main results on query rewriting for answering queries using views are surveyed in [11]. In contrast with query rewriting algorithms, views are not part of the input of view selection, but are part of the output together with the rewritings. In particular, and following [21], our view selection algorithm generates rewritings while searching for candidate views. As for the rewritings themselves, view selection produces equivalent rewritings, as query rewriting does in the setting of query optimization, while query rewriting for data integration typically produces maximally-contained rewritings due to the incompleteness of the data sources.

Materialized view selection has been intensely studied in relational databases [8] and data warehouses [12]. We used [21] as a starting point for our work, as it is one of the prevalent works in the area and the closest to our problem definition and query language. However, in [21] the restriction that *no relation may appear twice in a workload query* is imposed, under which view equivalence can be tested in PTIME. This simplification is incompatible with RDF queries, which repeatedly use the triple table. In our context, determining view equivalence (needed for VF and for the search strategies) is NP-complete [7]. This, along with the typically bigger size of RDF queries compared to the relational ones (since only one table with three attributes is used), increase the complexity of the problem even more. Hence, the strategies presented in [21] are not effective in our context. We innovate over [21] by proposing new search strategies and heuristics, which, as demonstrated in Section 6, do not suffer from memory limitations and lead to the selection of efficient views, even if we limit the time of the search. Furthermore, there are some differences between our transitions and those in [21], due to the differences between their SQL-like language and our Datalog formalism (for more details see [25]).

Multi-query optimization [28] and partial view materialization [29] are also related works. Unlike our approach, none of them aims to completely rewrite the queries using the views. In [28], common query subexpressions among the queries are recognized to be materialized. Views with disjunctions are supported, which we also plan to do as future work. In [29] views are only partially materialized and their content is adjusted as the queries change, which is another difference with our work (we consider static queries).

Query reformulation (a.k.a. unfolding) is directly related to query answering under constraints interpreted in an open-world assumption (e.g., [19]), i.e., when constraints are used as deductive rules. In particular, our query reformulation algorithm builds on those in the literature considering the so-called *Description Logic (DL) fragment* of RDF [3, 5], i.e., description logic constraints. This fragment corresponds to RDF databases without blank nodes that are made of an RDFS, called a Tbox, and a dataset made of assertions for classes and properties in the RDFS, called an Abox, i.e., well-formed triples of the form $(s, \text{rdf:type}, c)$ or $(s, p, o)$, where $c$ is a class and $p$ a property of the RDFS. Lastly, the RDF entailment rules considered are only those dedicated to an RDFS (see Section 4.1). Reformulation algorithms for the DL fragment of RDF actually reformulate queries from a strictly less expressive language than our RDF queries. They only support atoms in which the class or the property is specified, i.e., they do not support atoms of the form $t(s, \text{rdf:type}, X)$ or $t(s, X, o)$ with $X$ a variable. To overcome this, our reformulation algorithm extends the state of the art to our RDF queries, i.e., the BGP of SPARQL.

An early version of this work was demonstrated in [10].

## 8. CONCLUSION AND FUTURE WORK

We considered the setting of a Semantic Web database, including both explicit data encoded in RDF triples, and implicit data, derived from the RDF entailment rules [26]. Implicit data is important since correctly evaluating a query against an RDF database also requires taking it into account. In this context, we have addressed the problem of efficiently recommending a set of views to materialize, minimizing a combination of query evaluation, view storage and view maintenance costs. Starting from an existing relational approach, we have proposed new search algorithms and shown that they scale to large query workloads, for which previous search algorithms fail. Our view selection approach can be used as well with a saturated RDF database (where all implicit triples are added explicitly to the data), or with a non-saturated one (when queries need to be reformulated to reflect implicit triples). We have proposed a new algorithm for reformulating queries based on an RDF Schema, as well as a novel post-reformulation method for taking into account implicit triples in a query reformulation context. Post-reformulation can be much more efficient than naïve pre-reformulation, due to the high complexity of view search in the number of queries.

As future work, we consider parallelizing our view search algorithms by identifying workload queries that do not have many commonalities and running the search in parallel for each group. We also consider extending our query and view language, as well as adapting our approach to dynamic query workloads.